\documentclass[10pt]{bmc_article}

\usepackage{cite} 

\usepackage{url}  

\usepackage{ifthen}  

\usepackage{multicol}   

\usepackage{inputenc} 

\usepackage{amsmath,epsfig,float}

\newcommand{\etc}{\it etc.\rm}

\newcommand{\ie}{\it i.e. \rm}
\newcommand{\eg}{\it e.g. \rm }

\newcommand{\etal}{\it et al. \rm }
\newcommand{\celeg}{\it C.elegans\rm}
\newcommand{\scere}{\it S.cerevisiae\rm}
\newboolean{publ}

\newenvironment{bmcformat}{\begin{raggedright}\baselineskip20pt\sloppy\setboolean{publ}{false}}{\end{raggedright}\baselineskip20pt\sloppy}

\begin{document}

\begin{bmcformat}

\title{Comparative Analysis of the {\it Saccharomyces cerevisiae} and {\it Caenorhabditis elegans} Protein Interaction Networks}

\author{
        Ino Agrafioti$^1$
        \email{Ino Agrafioti - ino.agrafioti@imperial.ac.uk}%
        \and
        Jonathan Swire$^1$
        \email{Jonathan Swire - j@robberfly.com}
          \and
        James Abbott$^2$
        \email{James Abbott - j.abbott@imperial.ac.uk}
        \and
        Derek Huntley$^2$
        \email{Derek Huntley - d.huntley@imperial.ac.uk}
        \and
        Sarah Butcher$^2$
        \email{Sarah Butcher - s.butcher@imperial.ac.uk}
        and 
        Michael P.H. Stumpf\correspondingauthor$^1$
        \email{Michael P.H. Stumpf\correspondingauthor - m.stumpf@imperial.ac.uk}
      }


\address{%
    \iid(1)Department of Biological Sciences, Imperial College London,
        Wolfson Building, SW7 2AZ London, UK
    \iid(2)Bioinformatics Support Service, Imperial College London, Wolfson Building, SW7 2 AZ London, UK
    }%

\maketitle

\begin{abstract}


        \paragraph*{Background:} 

Protein interaction networks aim to summarize the complex interplay of proteins in an organism. Early studies suggested that the position of a protein in the network determines its evolutionary rate but there has been considerable disagreement as to what extent other factors, such as protein abundance, modify this reported dependence. 

 \paragraph*{Methods:} We compare the genomes of {\it Saccharomyces cerevisiae} and {\it Caenorhabditis elegans} with those of closely related species to elucidate the recent evolutionary history of their respective protein interaction networks. Interaction and expression data are studied in the light of a detailed phylogenetic analysis. The underlying network structure is incorporated explicitly into the statistical analysis. 

\paragraph*{Results:} The increased phylogenetic resolution, paired with high-quality interaction data, allows us to resolve the way in which protein interaction network structure and abundance of proteins affect the evolutionary rate. We find that expression levels are better predictors of the evolutionary rate than a protein's connectivity. Detailed analysis of the two organisms also shows that the evolutionary rates of interacting proteins are not sufficiently similar to be mutually predictive. 

\paragraph*{Conclusions:} It appears that meaningful inferences about the evolution of protein interaction networks require comparative analysis of reasonably closely related species. The signature of protein evolution is shaped by a protein's abundance in the organism and its function and the biological process it is involved in. Its position in the interaction networks and its connectivity may modulate this but they appear to have only minor influence on a protein's evolutionary rate. 

\end{abstract}

\ifthenelse{\boolean{publ}}{\begin{multicols}{2}}{}









\section*{Background}

Studies of the evolutionary history of protein interaction network (PIN) data have produced an almost bewildering range of (partially) contradictory results\cite{Wagner2001,Fraser2002,Jordan2003,Fraser2003,Bloom2003,Grigoriev2003,Han2004,Bader2004,Bloom2004,Fraser2004,Hahn2004}. While PIN data is notoriously prone to false positive and negative results\cite{Mering2002,Bloom2003}, reasons for disagreements are probably more diverse than just the quality of the interaction data. Failure to account for protein abundance --- as measured by gene expression levels, or by proxy, the codon-adaptation index --- has been criticized\cite{Jordan2003};  the choice of species for comparative analysis will also affect any evolutionary inferences as shown by Hahn \etal \cite{Hahn2004}. This may either be due to loss of power (\eg fewer reliably identified orthologues between more distantly related species) or to differences in underlying PINs in distantly related species. Below, for example, we will show that results obtained from a comparison between the two hemiascomycetes {\it Saccharomyces cerevisiae} and {\it Candida albicans} differ considerably from those obtained using a distant {\it S.cerevisiae}---{\it Caenorhabditis elegans} comparison. Finally, it has recently been shown that many studies may have suffered from the fact that present network data, and this is in particular true for PINs, are random samples from much larger networks. Unless these subnets are adequate representations of the overall network, their structural properties (such as node connectivity) may differ quite substantially from that of the nodes in the global network. This is, for example, the case for so-called scale-free network models\cite{Albert2002}. 

\par

Moreover, many studies have ignored the underlying network structure\cite{Maslov2002} in the statistical analysis. The network, however, introduces dependencies between connected proteins which should not be ignored. Fraser \etal \cite{Fraser2002} for example find that (i) there is a negative correlation between a protein's evolutionary rate and its connectivity $k$ (the number of its interactions), (ii) connected proteins have positively correlated evolutionary rates, and (iii) connected proteins do not have correlated connectivities. All three statements cannot, of course, be strongly true simultaneously. Here we observe only relatively weak ---though statistically significant--- correlations between connectivity and evolutionary rate. We will argue that in a regression framework\cite{Davison2003} some of these quantities contain very little information indeed about the corresponding properties of their interaction partners. Furthermore, we will demonstrate that when analyzing network data the network structure must be included into the analysis from the outset. 

\par

Here we will first perform an evolutionary analysis of the yeast and nematode PIN data available in the DIP data base (\url{http://dip.doe-mb.ucla.edu}), a hand-curated data-set combining information from a wide range of sources, followed by a comparison of the two data sets. When making comparisons between yeast species and between nematode species, we use only a single PIN data-set ---for \scere\ and \celeg, respectively--- and take comfort from the observation of Hahn \etal \cite{Hahn2004} who find that  evolutionary analysis involving closely related reference taxa produces consistent results. Previously, topological comparisons of biological network data from different species have been made\cite{Milo2004} but here we focus on shared evolutionary characteristics of PINs in the two species. We would expect at least some level of similarity of biological networks between species; but the more distantly related two organisms are, the more changes can have accumulated in their respective molecular networks. Thus the depth of the phylogeny can affect the evolutionary analysis of PINs; it is, for example, unlikely that PINs have been conserved over large evolutionary time-scales.

\par




\section*{Results}

\subsection*{Evolutionary Analysis of the \scere\ {\bf PIN}}

For the evolutionary analysis of the yeast PIN we use a panel of related yeast species: {\it Saccharomyces mikatae}, {\it Saccharomyces bayanus}, {\it Saccharomyces castelii}, {\it Saccharomyces kluyverii}, {\it C.albicans} and {\it Schizosaccharomyces pombe} (see  Methods section); the evolutionary relationship between these species is shown in figure 1. We thus focus on relatively recent evolutionary change which allows us to study the effects of the network structure on the rate of evolution more directly than \eg distant comparisons of \scere\ and \celeg, which may, after all, have different PINs.

\subsubsection*{Connectivity, expression and evolutionary rates in the \scere PIN}

For most protein sequences we have not been able to identify orthologues in all yeast species used in this analysis. We therefore defined the averaged relative evolutionary rate $R$ (see Eqn. (\ref{avgrate}) in the {\it Methods} section) which allows us to make comparisons for 4124 out of the 4773 yeast genes for which we have interaction data. 

\par

In figure 2 we show the dependence between inferred evolutionary rates and connectivities and expression levels. Our comparative analysis found statistically significant, though small, negative correlation, measured by Kendall's $\tau$, between estimated evolutionary rates and a protein's number of interactions. In table 1 and figure 3 we observe that comparisons with all species support this notion We furthermore estimated approximate confidence intervals for $\tau$ from 1000 bootstrap replicates \cite{Efron1998} (shown in table 1).

\par

Observed negative correlations between estimated evolutionary rates and the expression level ---which have been reported previously by Pal \etal \cite{Pal2001}--- are more pronounced. Equally, $k$, the number of a protein's interactions, and expression levels are also correlated ($\tau=0.09$). There has been considerable controversy as to whether the effect of a protein's connectivity can be studied independently of expression levels (see \eg  \cite{Jordan2003,Fraser2003}). The observed values of $\tau$ suggest that expression levels are better predictors of the evolutionary rate than are connectivities. Calculating partial rank correlation coefficients, $\tau_p$,  provides further evidence for this: correcting for expression reduces the correlation between  the evolutionary rate $R$ (or any of the individual rates) and the number of interactions, as is apparent from figure 3.  As the phylogenetic distance between species increases, the negative partial correlation between evolutionary rate and connectivity decreases compared to the uncorrected rank correlation measure $\tau$.  

\par

In the supplementary tables S1-S3 we show the evolutionary rates for the different functional categories, processes and cellular compartments (taken from {\it Gene ontology} (GO),  \url{www.geneontology.org}). 
Interestingly, once the effects of expression and protein function on the estimated evolutionary rate are taken into account the dependence of the latter on connectivity in a generalized linear regression model\cite{Davison2003} (where we log-transformed the expression level to obtain an approximately normal distribution)  is considerably reduced. This can be assessed formally using the Akaike information criterion (AIC)\cite{Burnham1998} on the sub-models where one of the terms has been dropped (see methods). For the full model we obtain AIC=-407.4. Dropping expression from the model results in AIC=-196.9, indicating that a substantial amount of information about the evolutionary rate is contained in the expression levels. Dropping the other terms individually while retaining the rest results in: AIC=-392.9 if the connectivity is dropped from the statistical model, and AIC=-352.6 (process), -250.1 (function) and -392.7 (compartment). We thus have the following order of statistically inferred impact on the evolutionary rate(with a slight abuse of the notation): expression$>$function$>$process$>$connectivity$\approx$compartment.  Using the rates obtained from comparisons with the individual species results in the same ordering.

\subsubsection*{Evolution of interacting proteins in \scere}

 So far we have treated nodes/proteins as independent (using only their connectivities in the analysis) but we will now consider the extent to which interactions introduce dependencies into the data. It is intuitively plausible that interacting proteins have similar evolutionary rates, and this has indeed been reported by Fraser \etal \cite{Williams2000,Fraser2002} and studied by others, too, \eg \cite{Jordan2003,Hahn2004}. Just like them we find that evolutionary rate decreases with connectivity; we also observe that the connectivities of interacting proteins are anti-correlated in yeast ($\tau\approx-0.03$ with $p<10^{-8}$). This is well explained from the statistical theory  of networks\cite{Albert2002,Newman2003b}, as well as structural analyses of PIN data, where it is found that highly connected proteins form hubs which connect sparsely connected proteins.
\par
Taken together this would mean that the evolutionary rates of connected proteins should also be anti-correlated. This is, however, not the case when we look at the yeast PIN, where we   find that evolutionary rates of interacting proteins are positively correlated as measured by Kendall's $\tau$. The correlations we observe are only relatively weak (even though they are significant) $\tau\approx0.05-0.10$ with $p<10^{-8}$. In figure 4 we show the distribution of the $\tau$ rank correlation under the correct network Null model (see methods) for rates, expression levels and connectivities of interacting proteins. The observed value always lies outside the distribution of the expected values. Also shown in the figure are the probabilities that two interacting proteins have identical GO-classifications for function, process and cellular compartment, respectively. Again the observed probabilities lie outside the distribution under the Null model. 
\par
Correlation, even partial correlation, may, however, be an inadequate statistical measure if the data is structured (as in a network); one should then rather focus on the power of a factor such as expression level or connectivity to predict evolutionary rates. We assess this formally through the use of statistical regression models which describe the evolutionary rate of one protein as a function of the rate of its interacting partner, as well as of its expression level, number of interactions, function, process and cell compartment. The AIC, which for the full model is AIC=-2397.6, allows us to order the factors by the information they contain about a protein's evolutionary rate. The order (and the respective AIC value on dropping the factor from the model) is as follows: Expression (AIC=-1399.6), function (AIC=-1445.9), process (AIC=-1956.6), cellular compartment (AIC=-2226.6), connectivity (AIC=-2316.8), and the rate of one of its interaction partners (AIC=-2397.0). Note that, measured by the AIC, the evolutionary rate of an interaction partner provides virtually no additional information about a protein's own evolutionary rate, once the protein's own expression level, function and process have been taken into account.

\par

Thus, in summary, we observe that the evolutionary rate of yeast proteins is inversely related both to their connectivity in the PIN and to their expression levels; with expression levels having a greater impact on a protein's evolutionary rate than connectivities. 
Finally, while there is statistically significant correlation between the rates of interacting proteins, the rate of one interaction partner carries very little information about the rate of the other protein if other factors are taken into account.

\subsection*{Evolutionary Analysis of the \celeg\ PIN}

In the evolutionary analysis of \celeg\ we use {\it C.briggsae}, the only other congeneric nematode for which high quality whole-genome data is available.  Since nematodes are multicellular, care has to be taken when analysing the effects of gene expression on evolutionary rate, as expression levels will vary considerably between tissues and, indeed, between different stages of the nematode life cycle. Because codon usage bias as a selective response increasing translational efficiency should be driven by the overall expression level of a protein integrated over both tissue and time, the codon-adaptation index (CAI; see Methods and \cite{Sharp1987}) can serve as a meaningful averaged quantity reflecting overall integrated expression levels better than a direct measurement of mRNA expression level data obtained from any single tissue type.

\subsubsection*{Connectivity, expression and evolutionary rates in the \celeg\ PIN}

The correlation of evolutionary rate and connectivity is somewhat reduced compared to \scere\ with a point estimate of $\tau=-0.05$ with a 95\% bootstrap CI of $[-0.097,-0.017]$. Anti-correlation between the CAI measure of expression and evolutionary rates is again much more pronounced with $\tau\approx -0.30$ and approximate bootstrap CIs of $[-0.333,-0.264]$. The resulting scatter plots of rate vs. connectivity and rate vs. CAI are shown in figure 5.
 \par
Partial correlation coefficients again show that the influence of expression is greater than that of connectivity: $\tau_p\approx-0.03$ for the partial correlation measure between rate and connectivity, while $\tau_p\approx -0.30$ if the correlation between expression (CAI) and rate is corrected for connectivity.  This is confirmed by performing an ANOVA \cite{Rice1995} on the regression between rate, CAI and connectivity where no significant correlation can be found between rate and connectivity ($p\approx0.62$). Generalized linear regression modelling shows that measured by the AIC a model in which the rate depends only on the CAI but not on the connectivity (AIC=-660.5) is more powerful than a model in which the rate depends on both connectivity and CAI (AIC=-618.4). In the absence of extensive GO data we find that the CAI is the only statistically significant predictor for a protein's evolutionary rate.

\subsubsection*{Evolution of interacting proteins in \celeg}

Comparing properties of interacting proteins we again find a negative correlation between their respective connectivities ($\tau=-0.07$) and a weaker correlation between their evolutionary rates  ($\tau=0.03$). The corresponding 95\% bootstrap CI for $\tau$ does, however, include 0 and negative values; thus there is no statistical basis for concluding that evolutionary rates of interacting proteins are correlated in \celeg\ even if we consider only the rank correlation measure.  In figure 6 the distribution of $\tau$ under the correct Null model (see methods) confirms this result as the observed correlation between the evolutionary rates of interacting proteins falls into the 95\% confidence interval obtained from the Null model. Expression levels are, however, significantly correlated and connectivities remain significantly anti-correlated. Regression models, equivalent to those performed for yeast, confirm the negligible information a protein's evolutionary rate contains about the evolutionary rate of an interacting protein. 
\par
In summary, for \celeg\ we find that expression, even if measured indirectly through the CAI, is a better predictor about a protein's evolutionary rate than connectivity and GO classifications. The evolutionary rates of connected proteins do not appear to be correlated.

\subsection*{Comparing the PINs of \scere\ \bf and \celeg}

It is instructive to compare the PINs of the two model organisms, yeast and worm, directly. We have therefore used our earlier approach of identifying and analysing orthologues to the yeast and nematode PIN data. While we are, of course, aware that this may be problematic given the two or three billion years of evolutionary history separating the two organisms,  it should serve as a useful illustration of the amount of information one model-organism is likely to provide about another (including, of course, humans). 
\par
Using this approach we found a total of 524 pairs of orthologues. These we aligned and from the alignments we estimated evolutionary rates. For all of these proteins we have PIN data and for most we also have information about their expression levels in the two species. The results are summarized in tables 2 and 3. Although they essentially agree with the earlier results, they do suggest that the choice of species used for inferring the evolutionary rate can influence the analysis. For example, the partial correlation between interaction and evolutionary rate (calculated directly from the \scere---\celeg\ amino acid sequence comparison) accounting for expression is much less reduced compared with the simple correlation coefficient ($\tau_p=-0.20$ in \scere, and $\tau_p=-0.10$ in \celeg) than when  evolutionary rates are calculated using more closely related target species. Over long evolutionary distances it appears as if connectivity and expression level act almost independently. However, the more reliable comparisons of the previous section suggest that this is not the case.
 \par
Comparing properties of orthologous proteins we find that their expression levels (using the CAI as a proxy in \celeg) show the strongest correlation while their respective PIN connectivities show the lowest value for Kendall's $\tau$ statistic. This may be due to the noise in the PIN data or the incomplete nature of present PIN data sets. We expect that the relatively small proportion of \celeg\ proteins included in the DIP database will also lead to an inaccurate representation of the \celeg\ PIN. 


\section*{Discussion}

There are considerable differences between the various published studies\cite{Fraser2002,Fraser2003,Jordan2003,Hahn2004}, both in terms of protein interaction data and phylogenetic comparisons. We therefore focus on closely related species for both \scere\ and \celeg\ in the evolutionary analysis, since we probably have to assume that the underlying PIN is relatively more conserved over short evolutionary distances. 
While we found some evidence that highly connected proteins evolve more slowly than sparsely connected proteins, (i) the negative correlation between rate and expression level is more pronounced, (ii) in \scere\ and \celeg\ connectivity turns out to be a worse statistical predictor of the evolutionary rate than expression. For \scere\ we also find that protein function and the principal biological process a protein is involved in have a greater impact on the evolutionary rate of a protein than its connectivity. 
\par
We believe that the importance of expression over connectivity in determining the evolutionary rate may be due to three factors. First, highly abundant genes are perhaps more visible to purifying selection\cite{Li1997}, which might tend to lower the rate at which they evolve. Second, and more importantly,  highly expressed genes, which are under selection for translational efficiency, use only a small subset of the cognate codons for a particular amino acid (this, incidentally, is exploited in the construction of the CAI), and because this subset is often the same even in phylogenetically remote organisms ---for example, for those amino acids encoded by nnU and nnC (\eg phenylalanine or cysteine), nnC is almost universally preferred--- the silent substitution rate is reduced. Third, the replacement substitution rate in highly expressed proteins should also be reduced for a similar reason to selection for translational efficiency at silent sites:  selection for more cheaply synthesised amino acids at replacement sites \cite{Akashi2002}. This can be shown to lead to the avoidance of amino acids which are metabolically expensive to synthesise at functionally-unconstrained sites in highly expressed proteins, which reduces the set of acceptable amino acids at such sites and thereby lowers the replacement substitution rate compared with that at functionally-unconstrained sites in low expression proteins\cite{Swire2003}.  
\par
We have also applied an improved resampling procedure to the analysis of correlation between rates and expression levels of connected proteins. In our analysis we treated properties of the network as a confounding variable and in addition to studying correlations we also show how informative properties of one protein are about properties of its interaction partners. We find that the correct procedure broadens the resampling confidence intervals but that expression levels of interacting proteins remain considerably closer than would be expected by chance. Conversely we found no evidence of a correlation in the evolutionary rate of interacting proteins in \celeg\ and only extremely weak evidence in \scere. 
Our results also suggest, that the evolution of interacting proteins is not as tightly correlated as some researchers have proposed. This level of disagreement may be caused by uncertainties in the data or the fact that subnets sampled from larger networks inaccurately reflect the properties of the true network \cite{Stumpf2005}.   %


\section*{Conclusions}

We believe that the effects of the network structure on the evolution of proteins, and vice versa, is much more subtle than has previously been suggested.  In the present dataset expression levels appear to have shaped a protein's evolutionary rate more than its connectivity. If we are happy to accept present PIN data with the necessary caution, then this observation is consistent with a scenario where expression levels are more conserved between species than are details of the interaction network. Nevertheless, we believe that it is important to consider the PIN and a protein's connectivity explicitly and from the outset in any statistical analysis as the underlying network appears to act as a confounding factor.


\section*{Methods}

  \subsection*{Data}

  \subsubsection*{Protein interaction data}

The names and sequences for proteins with known interactions in Saccharomyces cerevisiae and Caenorhabditis elegans were retrieved from the Database of Interacting Proteins (DIP) on the 5th of July (\url{http://dip.doe-mbi.ucla.edu/}).  The database mainly contains information extracted from the research literature, but recently the database was enriched with information obtained by analysing structures of protein complexes deposited in PDB\cite{Salwinski2004}. We have data for 4773 yeast proteins (comprising 15461 interactions) and 2386 nematode proteins (with 7221 interactions). While there have been recent attempts at quantifying levels of confidence in given protein interactions these generally lead to substantial decreases in sample size. For this reason we have therefore chosen to take the PIN data as it is deposited in the hand-curated DIP data-base (we have also performed analyses with such restricted subsets which agree with the results presented here).

\subsubsection*{Protein sequence data}
In addition to \scere\ sequences downloaded from the DIP, publicly available protein sequences of six other yeast species were investigated: {\it Saccharomyces pombe} (\url{ftp://ftp.sanger.ac.uk/pub/yeast/pombe/Protein_data/pompep}), {\it Candida albicans} (\url{ftp://cycle.stanford.edu/pub/projects/candida/}), {\it Saccharomyces mikatae, Saccharomyces bayanus, Saccharomyces kluyverii and Saccharomyces castelii} (the last four all from \url{ftp://genome-ftp.stanford.edu/pub/yeast/data_download/sequence/fungal_genomes/}). Genomic protein sequences for only one other Caenorhabditis species apart from \celeg\ are publicly available ({\it C.briggsae}); these were downloaded from \url{ftp://ftp.ensembl.org/pub/current_cbriggsae/data/fasta/pep/}. All sequence files were converted to searchable indexed databases; these are available from the authors.

\subsubsection*{Expression data}
\scere\ expression data came from Cho et al. \cite{Cho1998} who characterised all mRNA transcript levels during the cell cycle of \scere. mRNA levels were measured at 17 time points at 10 min intervals, covering nearly two full cell cycles. Thus, in the supplementary file of this paper, which is available online (\url{http://sgdlite.princeton.edu/download/yeast_datasets/expression/Cho_et_al_full_data.txt}), 
for each protein there were 17 different numbers, one for each time point. The mean of these 17 numbers was taken to produce one general time-averaged expression level for each protein.  
\par

\celeg\ is a multicellular organism in which different cells have different 
functions. This means that different proteins have different expression levels in different cells, which are present in different numbers, so taking a simple mean of the expression levels in a single cell type would be pointless. In addition, \celeg\ has a complex life-cycle, with different proteins being expressed in different stages of that cycle. Thus, an alternative way of calculating a single expression level for each protein had to be used. It has long been known that highly expressed genes tend to use only a limited number of codons thus displaying high codon bias. Sharp and Li\cite{Sharp1987} devised a measure for assessing the degree of deviation from a preferred pattern of usage estimated from the clustering of codon usage across proteins, which they called the Codon Adaptation Index (CAI).  We adopt this measure as out expression level proxy. 
  \par

  \subsection*{Methods}

  \subsubsection*{Phylogenetic Analysis\\}

The close relationship between the species considered here, apart from the distant comparison between \scere\ and \celeg, makes identification of orthologues relatively straightforward. Orthologous protein sequences were detected by reciprocal BLAST searches in the standard way. Multiple alignments of inferred orthologues were obtained using ClustalW.

  \par

Evolutionary rates were obtained using PAML (Phylogenetic Analysis by Maximum Likelihood, \url{http://abacus.gene.ucl.ac.uk/software/paml.html}). We used both the observed fraction of amino acid differences, referred to by $M1, B1, \ldots$ (where the letters refer to different species, see footnote to Table 1), and the distance related measure calculated from the trees inferred by PAML, referred to by $M2, B2, \ldots$. Both rates are highly correlated $\tau\approx 0.9$. In order to estimate the latter rate the phylogeny had to be reconstructed. Inferred phylogenies were assessed for their agreement with the commonly accepted family tree of yeast species (see figure 1; we found excellent agreement among the inferred trees assessed using the {\it clann} software package; \url{bioinf.may.ie/software/clann}) and the widely accepted phylogenies for the yeast and nematode species, which are shown in figure 1. For further analyses we chose to use the maximum likelihood rate. 

  \par

  \subsubsection*{Statistical Analysis\\}

  \par

In order to be able to compare evolutionary rates for as many proteins as possible we defined the averaged relative evolutionary rate of each protein $i$ via

  \begin{equation}
R_i = \frac{1}{\nu_i}\sum_{s\in \{\text{all species}\}} \frac{[\text{evol. rate of protein  $i$ form comparison with species $s$}]}{[\text{Avg. evol. rate in species  $s$ across all proteins}]}
  \label{avgrate}
  \end{equation}

where $\nu_i$ is the number of comparisons from which an evolutionary rate can be estimated. 
\par
We generally found that analysis of the dependence of $R=\{R_1,R_2,..R_n\}$ on the number of interactions \etc behaved similarly to analysis of the species specific rates. We used ANOVA \cite{Rice1995} and partial correlation coefficients to study the impact the different factors had on the evolutionary rates. All analysis was done using the R statistical environment and the NetZ package (available from the authors).
\par
In order to investigate the relative influence of the various factors (number of interactions, expression levels, GO-data) we used linear and generalized linear regression modelling (implemented in R). The Akaike information criterion (AIC) \cite{Burnham1998,Davison2003} was used to distinguish among the different nested submodels of the full model. The model which has the smallest AIC (defined as $2 (-\text{lk}(\theta)+2\nu)$ where $\text{lk}(\theta)$ is the log-likelihood of a ---potentially vector-valued--- parameter $\theta$, and $\nu$ is the number of parameters) is the model which offers the best (in an information sense\cite{Burnham1998}) description of the data. The full model included the number of interactions, expression levels (or CAI in the case of \celeg) and GO-data as explanatory variables. When comparing evolutionary rates of interacting proteins the evolutionary rate at the interacting protein was added as an explanatory variable. The AIC (and related approaches \cite{Burnham1998}) aims to identify a statistical model that offers the most efficient description of the data (in an information theoretic sense) from a set of trial models. 
\par
We explicitly incorporated the network structure into the statistical analysis. This is necessary if there is reason to believe that properties of the network may determine aspects of the evolutionary history, for example when we want to test if the evolutionary rates of interacting proteins are correlated. Here we use resampling or bootstrap procedures \cite{Efron1998} to determine if properties (\eg expression levels, rates, connectivities) of interacting proteins are more similar than would be expected to occur by chance. If instead we had paired proteins completely at random we would potentially have masked confounding effects due to the network (for example if expression depends strongly on a protein's network properties, \ie connectivity). In our network-aware resampling procedure we therefore pick each protein with a probability that is proportional to its number of interaction partners. Each bootstrap replicate is thus also a sample with the correct nodal properties and (statistically) the same degree distribution as the true network. In the structural analysis of networks the need to account for network properties in the construction of the correct Null model has long been realized \cite{Albert2002,Milo2004} but this is, to our knowledge, the first time that such a topologically correct Null model has been applied to the evolutionary analysis of network data. As an illustration the figure in this  section shows the bootstrap distribution of correlation coefficients of expression levels in yeast for the correct Null model and for the model where proteins are paired completely at random. Ignoring the correlation of the data introduced by the underlying network structure reduces the bootstrap confidence intervals considerably (we find that the two-sided 95\% CIs are reduced by approximately 20\% compared to the network aware bootstrap replicate). This mirrors the effects observed in population and evolutionary genetics where the underlying genealogy/phylogeny increases the CIs compared to the case of truly independent observations.

\par
Routines used to perform the statistical analysis of the network data are collected in the NetZ package which can be obtained from the corresponding author.


\section*{Authors' contributions}

IA collected the data, performed the phylogenetic analysis, helped with the statistical analysis and with writing the manuscript. JS calculated the codon adaptation index, helped with the evolutionary and statistical analysis and writing the paper. JA, DH and SB helped with data retrieval and the construction of a pipeline for the evolutionary analysis of the PIN data and writing the manuscript. MPHS devised the study, performed the statistical analysis and wrote the manuscript. All authors approved the final version of the manuscript.
\section*{Acknowledgements}
\ifthenelse{\boolean{publ}}{\small}{}
IA thanks the Wellcome Trust for a PhD studentship in Bioinformatics. MPHS gratefully acknowledges a Wellcome Trust Research fellowship. This research was also funded by the Royal Society and EMBO. We thank Eric de Silva and Piers Ingram, Bob May and Carsten Wiuf for helpful discussions.


\begin{thebibliography}{10}
\providecommand{\url}[1]{[#1]}
\providecommand{\urlprefix}{}

\bibitem{Wagner2001}
Wagner A: \textbf{The yeast protein interaction network evolves rapidly and
  contains few redundant duplicate genes}. \emph{MOLECULAR BIOLOGY AND
  EVOLUTION} 2001, \textbf{18}(7):1283--1292.

\bibitem{Fraser2002}
Fraser HB, Hirsh AE, Steinmetz LM, Scharfe C, Feldman MW: \textbf{Evolutionary
  rate in the protein interaction network.} \emph{Science} 2002,
  \textbf{296}(5568):750--2,
  \urlprefix\url{[http://dx.doi.org/10.1126/science.1068696]}.

\bibitem{Jordan2003}
Jordan IK, Wolf YI, Koonin EV: \textbf{No simple dependence between protein
  evolution rate and the number of protein-protein interactions: only the most
  prolific interactors tend to evolve slowly.} \emph{BMC Evol Biol} 2003,
  \textbf{3}:1.

\bibitem{Fraser2003}
Fraser H, Wall D, Hirsh A: \textbf{A simple dependence between protein
  evolution rate and the number of protein-protein interactions}. \emph{BMC
  Evolutionary Biology} 2003, \textbf{3}:11,
  \urlprefix\url{[http://www.biomedcentral.com/1471-2148/3/11]}.

\bibitem{Bloom2003}
Bloom J, Adami C: \textbf{Apparent dependence of protein evolutionary rate on
  number of interactions is linked to biases in protein interactions
  data sets}. \emph{BMC Evolutionary Biology} 2003, \textbf{3}:21,
  \urlprefix\url{[http://www.biomedcentral.com/1471-2148/3/21]}.

\bibitem{Grigoriev2003}
Grigoriev A: \textbf{On the number of protein-protein interactions in the yeast
  proteome.} \emph{Nucleic Acids Res} 2003, \textbf{31}(14):4157--61.

\bibitem{Han2004}
Han J, Bertin N, Hao T, Goldberg D, Berriz G, Zhang L, Dupuy D, Walhout A,
  Cusick M, Roth F, Vidal M: \textbf{Evidence for dynamically organized
  modularity in the yeast protein-protein interaction network}. \emph{NATURE}
  2004, \textbf{430}(6995):88--93.

\bibitem{Bader2004}
Bader JS, Chaudhuri A, Rothberg JM, Chant J: \textbf{Gaining confidence in
  high-throughput protein interaction networks.} \emph{Nat Biotechnol} 2004,
  \textbf{22}:78--85, \urlprefix\url{[http://dx.doi.org/10.1038/nbt924]}.

\bibitem{Bloom2004}
Bloom J, Adami C: \textbf{Evolutionary rate depends on number of
  protein-protein interactions independently of gene expression level:
  Response}. \emph{BMC Evolutionary Biology} 2004, \textbf{4}:14,
  \urlprefix\url{[http://www.biomedcentral.com/1471-2148/4/14]}.

\bibitem{Fraser2004}
Fraser H, Hirsh A: \textbf{Evolutionary rate depends on number of
  protein-protein interactions independently of gene expression level}.
  \emph{BMC Evolutionary Biology} 2004, \textbf{4}:13,
  \urlprefix\url{[http://www.biomedcentral.com/1471-2148/4/13]}.

\bibitem{Hahn2004}
Hahn MW, Conant GC, Wagner A: \textbf{Molecular evolution in large genetic
  networks: does connectivity equal constraint?} \emph{J Mol Evol} 2004,
  \textbf{58}(2):203--11,
  \urlprefix\url{[http://dx.doi.org/10.1007/s00239-003-2544-0]}.

\bibitem{Mering2002}
von Mering C, Krause R, Snel B, Cornell M, Oliver SG, Fields S, Bork P:
  \textbf{Comparative assessment of large-scale data sets of protein-protein
  interactions.} \emph{Nature} 2002, \textbf{417}(6887):399--403,
  \urlprefix\url{[http://dx.doi.org/10.1038/nature750]}.

\bibitem{Albert2002}
Albert R, Barabasi A: \textbf{Statistical mechanics of complex networks}.
  \emph{REVIEWS OF MODERN PHYSICS} 2002, \textbf{74}:47--97.

\bibitem{Maslov2002}
Maslov S, Sneppen K: \textbf{Specificity and stability in topology of protein
  networks.} \emph{Science} 2002, \textbf{296}(5569):910--3,
  \urlprefix\url{[http://dx.doi.org/10.1126/science.1065103]}.

\bibitem{Davison2003}
Davison A: \emph{Statistical Methods}. Cambridge University Press 2003.

\bibitem{Milo2004}
Milo R, Itzkovitz S, Kashtan N, Levitt R, Shen-Orr S, Ayzenshtat I, Sheffer M,
  Alon U: \textbf{Superfamilies of evolved and designed networks}.
  \emph{SCIENCE} 2004, \textbf{303}(5663):1538--1542.

\bibitem{Efron1998}
Efron B, Tibshirani R: \emph{An introduction to the Bootstrap}. Chapman \&
  Hall/CRC 1998.

\bibitem{Pal2001}
Pal C, Papp B, Hurst L: \textbf{Highly expressed genes in yeast evolve slowly}.
  \emph{Genetics} 2001, \textbf{158}:927--931.

\bibitem{Burnham1998}
Burnham K, Anderson D: \emph{Model Selection and Multimodel Inference}.
  Springer 1998.

\bibitem{Williams2000}
Williams E, Hurst L: \textbf{The proteins of linked genes evolve at similar
  rates}. \emph{Nature} 2000, \textbf{407}:900--903.

\bibitem{Newman2003b}
Newman M: \textbf{The structure and function of complex networks}. \emph{SIAM
  REVIEW} 2003, \textbf{45}(2):167--256.

\bibitem{Sharp1987}
Sharp P, WH L: \textbf{The Codon Adaptation Index - a measure of directional
  synonymous codon usage bias, and its potential applications}. \emph{Nucl.Acid
  Res.} 1987, \textbf{15}:1281--1295.

\bibitem{Rice1995}
Rice JA: \emph{Mathematical Statistics and Data Analysis}. Duxbury Press 1995.

\bibitem{Li1997}
Li WH: \emph{Molecular Evolution}. Sunderland, MA: Sinauer Associates 1997.

\bibitem{Akashi2002}
Akashi H, Gojobri T: \textbf{Metabolic efficiency and amino acid composition in
  the proteomes of Escherichia coli and Bacilus subtilis}. \emph{PNAS} 2002,
  \textbf{99}:3695--3700.

\bibitem{Swire2003}
Swire J: \textbf{Selection on Cost as a Driver of Molecular Evolution}.
  \emph{Phd thesis}, Imperial College London 2003.

\bibitem{Stumpf2005}
Stumpf M, Wiuf C, May R: \textbf{Subnets of scale-free networks are not
  scale-free: the sampling properties of random networks}. \emph{PNAS} 2005.

\bibitem{Salwinski2004}
Salwinski L, Miller C, Smith A, Pettit F, Bowie J, Eisenberg D: \textbf{The
  database of interacting proteins: 2004 update}. \emph{Nucl.Acid Res.} 2004,
  \textbf{15}(3):D449--D451.

\bibitem{Cho1998}
Cho R, Campbell M, Winzeler E, Steinmetz L, Conway A, Wodicka L, Wolfsberg T,
  Gabrielian A, Landsman D, Lockhart D, Davies R: \textbf{A genome-wide
  transcriptional analysis of the mitotic cell cycle}. \emph{Mol.Cell} 1998,
  \textbf{2}:65--73.

\end{thebibliography}


\ifthenelse{\boolean{publ}}{\end{multicols}}{}

\section*{Figures}
\subsection*{Figure 1 - Phylogeny of the organisms used in the study}

  \begin{figure}[H]

  \epsfig{file=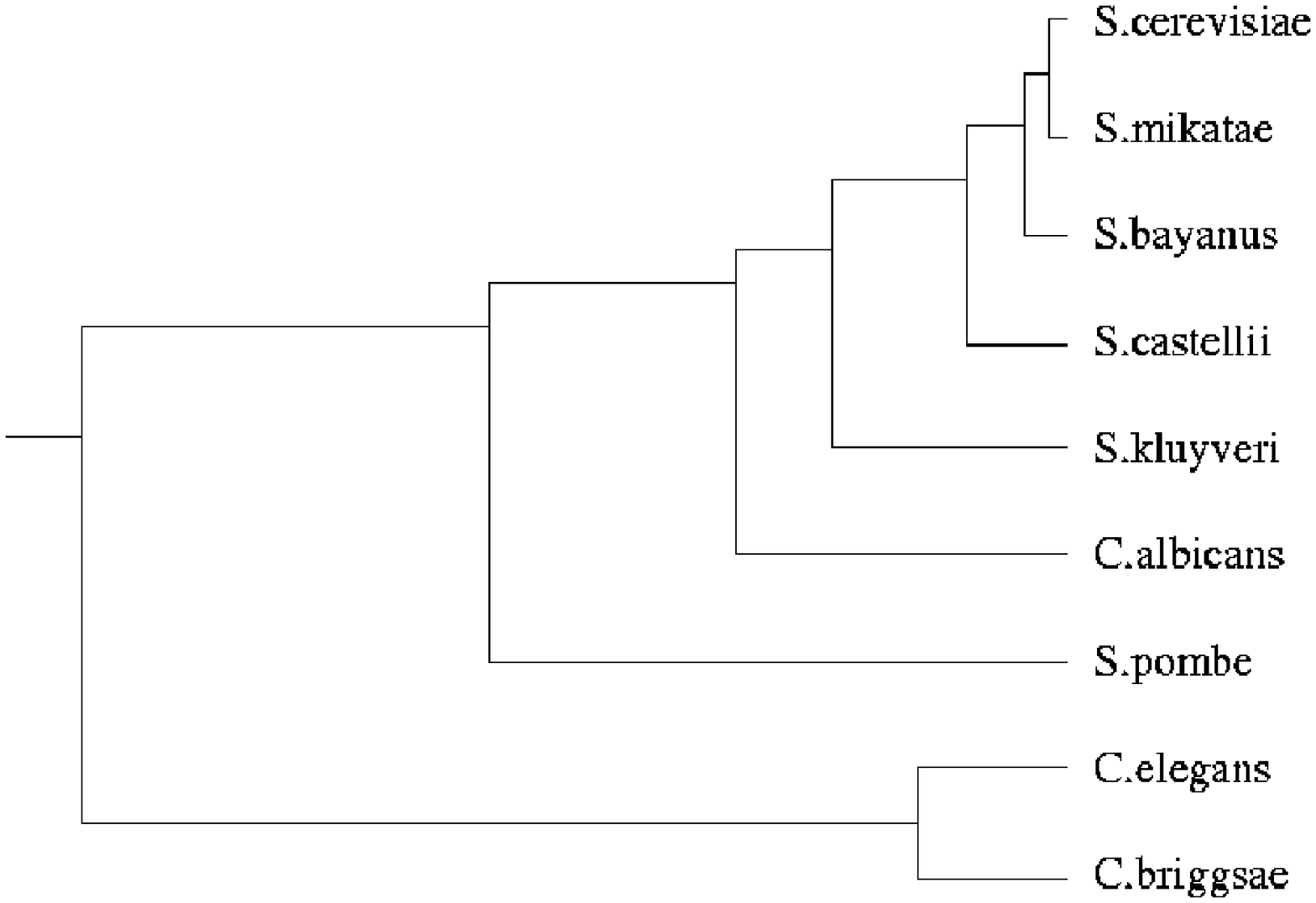,width=11cm}

  \end{figure}

The evolutionary relationship of the organisms used in this study. The last common ancestor of the ascomycetes in this phylogeny has been estimated to have lived approximately 330 million years ago. For the nematodes only two annotated genomes were available: their last common ancestor is believed to have lived approximately 100 million years ago.

  \subsection*{Figure 2 - Dependence of evolutionary rate in ascomycetes on the number of protein interactions and expression level}

  \begin{figure}[H]

  \epsfig{file=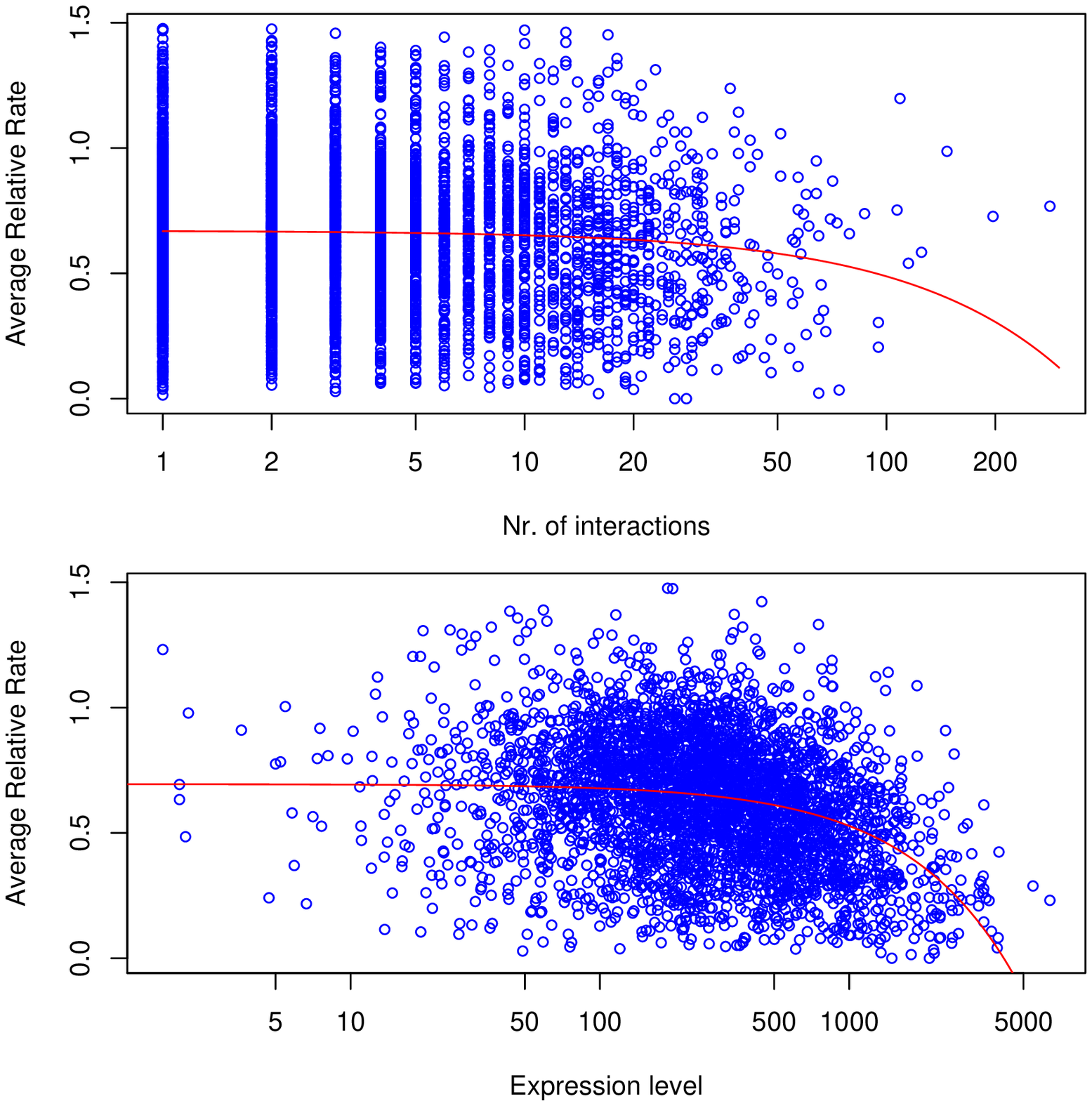,width=11cm}

  \end{figure}

The averaged relative rate $R$ decreases with increasing number of interaction partners ($\tau\approx-0.06$) and the expression level ($\tau\approx-0.23$). The 95\% bootstrap intervals for Kendall's $\tau$ values obtained from the six species comparisons are always negative (see table 1). The linear regression curves (red) appear concave on the log-transformed $x$-axis.

  \subsection*{Figure 3 - Correlation and partial correlation between evolutionary rate, number of  interactions and expression level}

  \begin{figure}[H]

  \epsfig{file=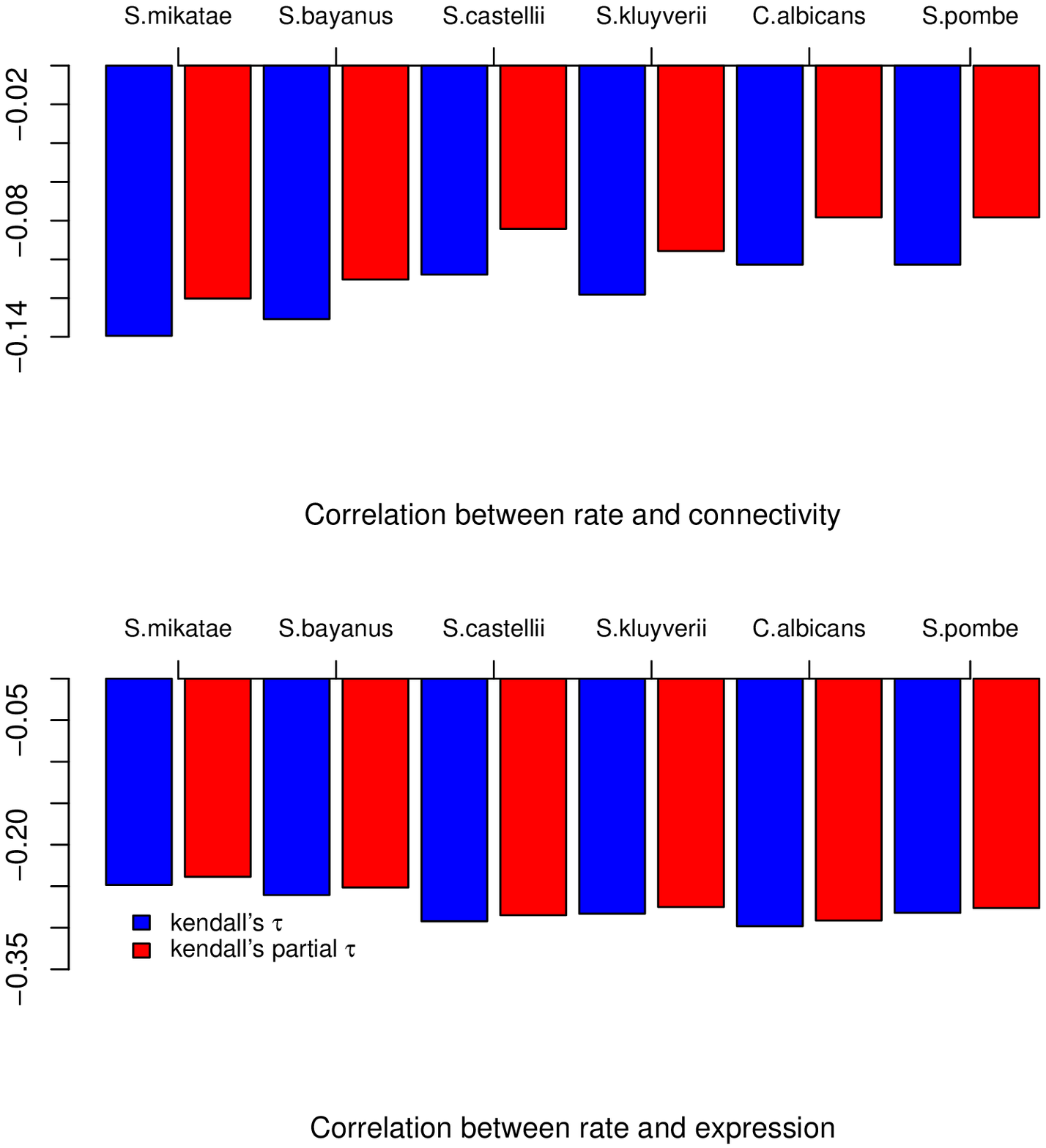,width=11cm}

  \end{figure}

Kendall's rank correlation (blue) and partial rank correlation coefficients (red) between $R$ and the number of interactions (correcting the partial $\tau$ for expression level) and expression (correcting for the number of interactions).

\subsection*{Figure 4 - Statistical dependencies of interacting proteins in \scere}

  \begin{figure}[H]

  \epsfig{file=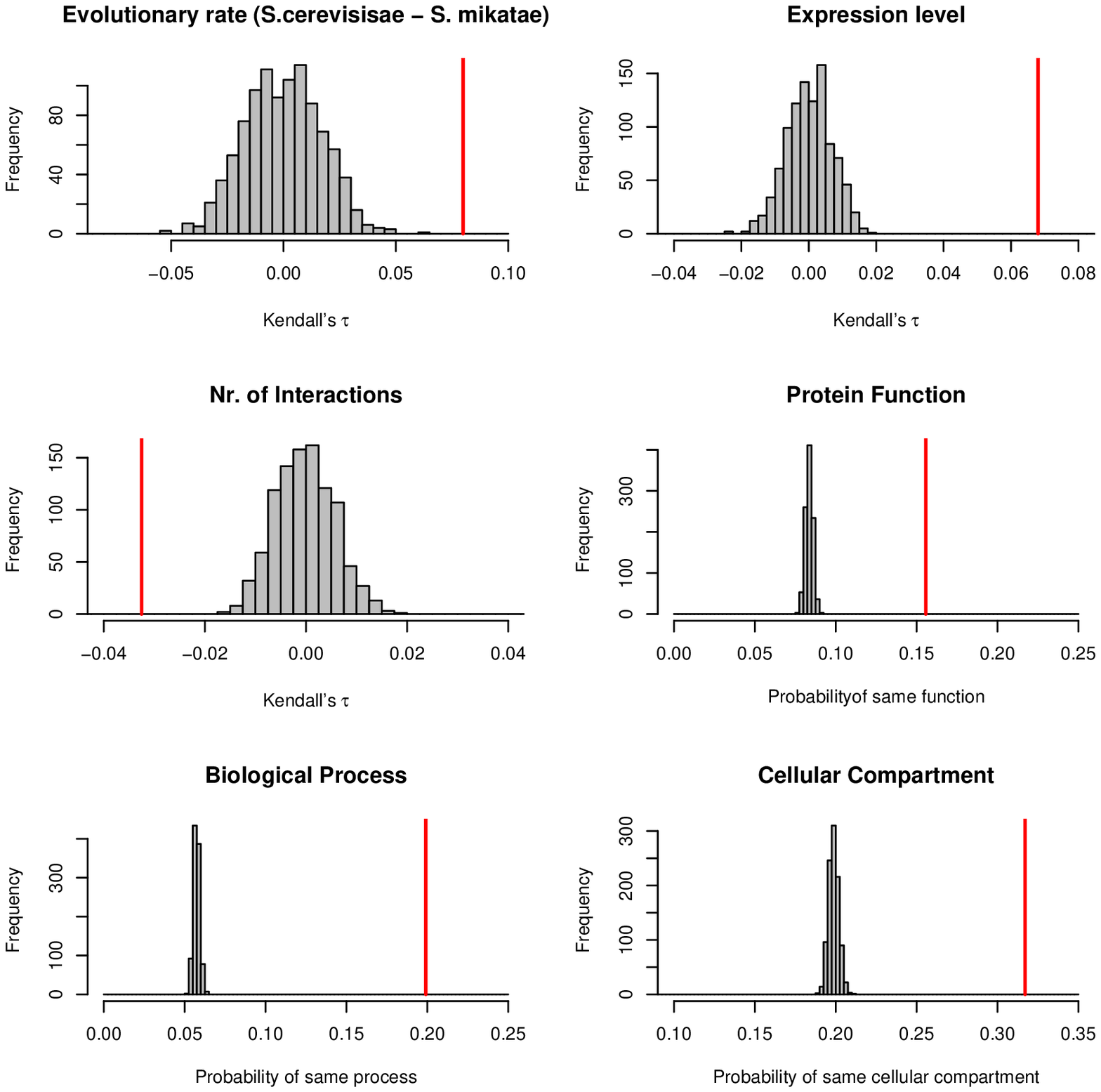,width=11cm}

  \end{figure}

Bootstrap distributions of Kendall's $\tau$ between evolutionary rates, expression levels and numbers of interactions and probabilities that protein function and the processes and cellular compartments by which proteins are classified are identical for a pair of interacting proteins. The grey histograms show the distribution of the statistics obtained from 1000 bootstrap replicates and the red vertical lines indicate the observed value. The bootstrap procedure was constrained such that each sample reproduced the degree distribution of the observed PIN.

\subsection*{Figure 5 - Dependence of evolutionary rate in nematodes on the number of protein interactions and CAI}

  \begin{figure}[H]

  \epsfig{file=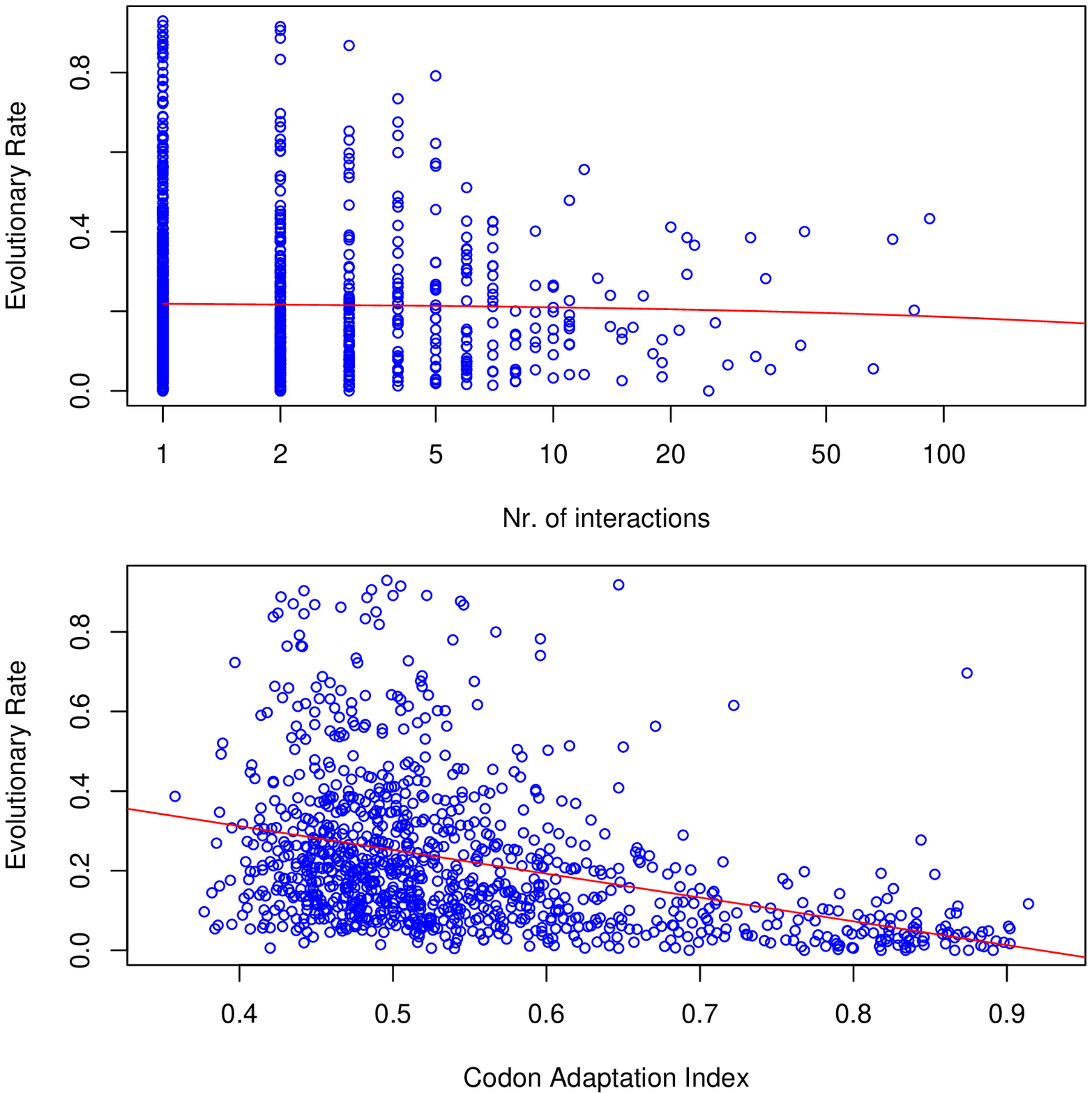,width=11cm}

  \end{figure}

The estimated evolutionary rate decreases with increasing number of interaction partners ($\tau\approx-0.03$) and the expression level ($\tau\approx-0.30$). We have again transformed the $x$-axis in the scatterplot of rate vs. connectivity which leads to the concave shape of the regression line (red).

\subsection*{Figure 6 - Statistical dependencies of interacting proteins in \celeg}

  \begin{figure}[H]

  \epsfig{file=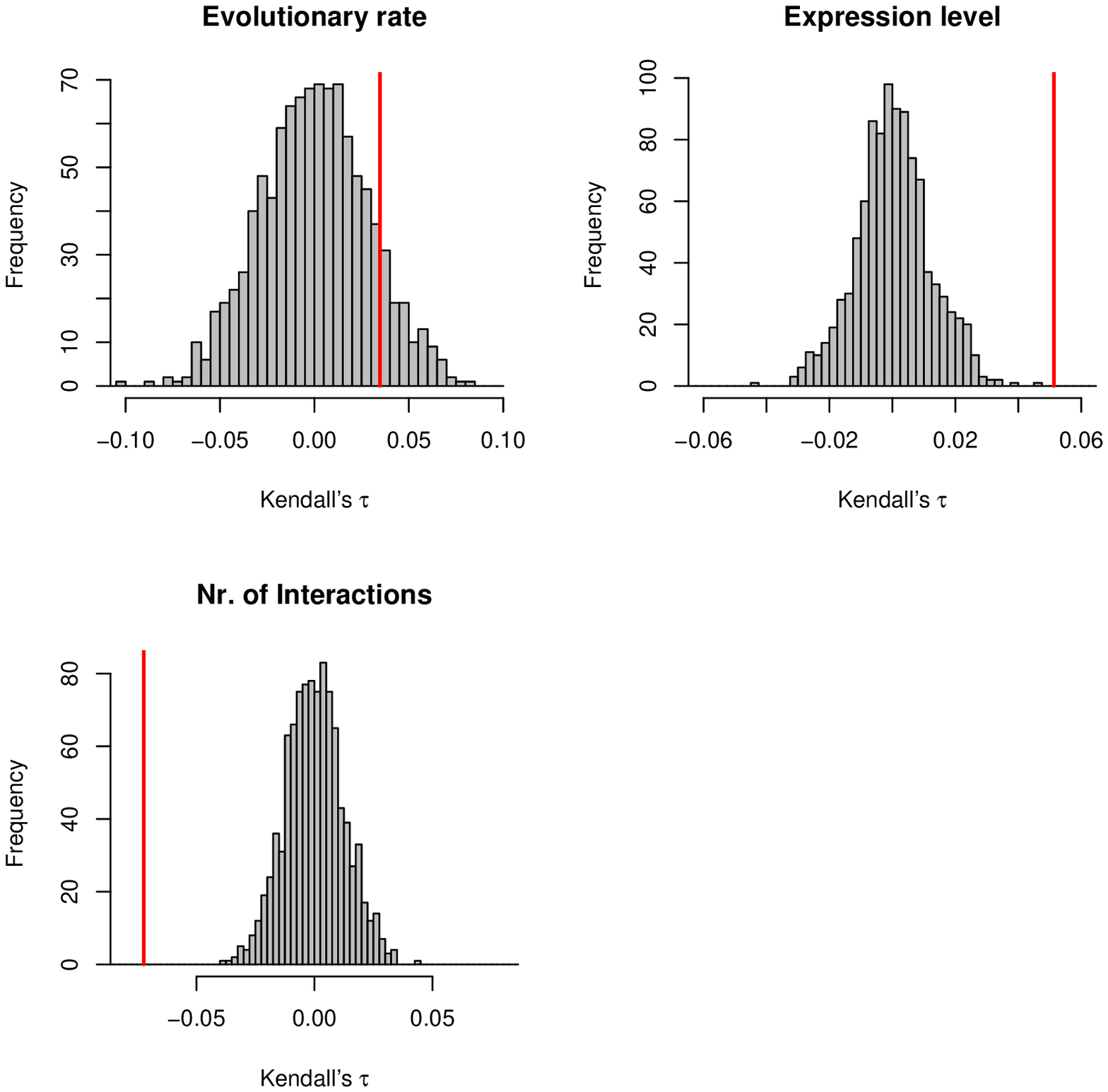,width=11cm}

  \end{figure}

Bootstrap distributions of Kendall's $\tau$ between evolutionary rates, expression levels and numbers of interactions. The grey histograms show the distribution of $\tau$ obtained in 1000 bootstrap replicates and the red lines indicate the observed value. The bootstrap procedure was constrained such that each sample reproduced the degree distribution of the observed PIN.

\subsection*{Figure Methods - Confidence intervals calculated with and without including the network structure.}

  \begin{figure}[H]

  \epsfig{file=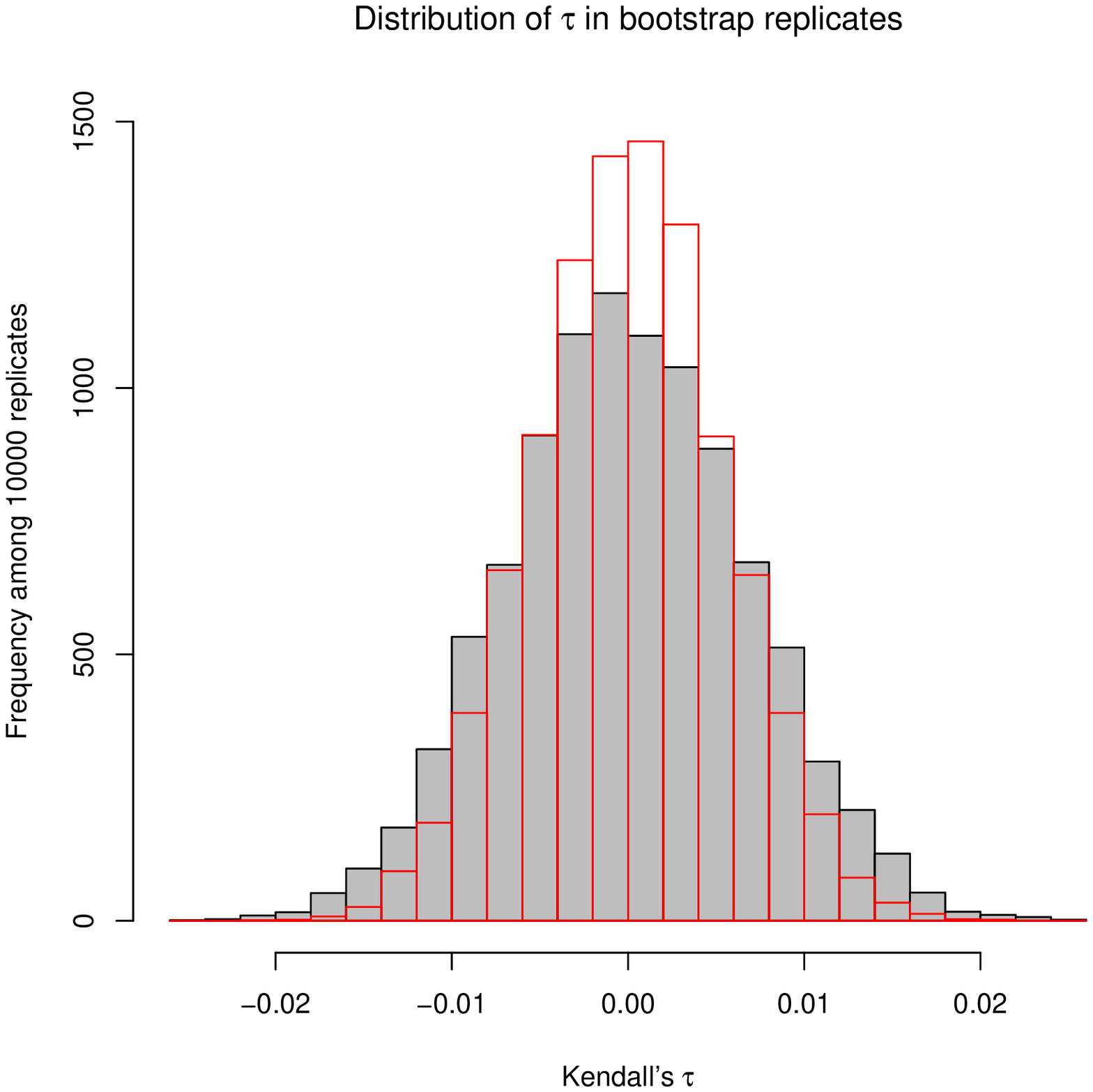,width=11cm}

  \end{figure}

Distribution of Kendall's $\tau$ measuring the expected correlation between the expression levels  of interacting proteins in \scere.    The grey distribution has been calculated under the correct empirical Null distribution, where pairs of interacting proteins are chosen such that the degree distribution of the re-sampled protein network is the same as that of the true network. The red bars indicate the distribution obtained under the conventional (and inadequate) Null model where pairs of proteins are chosen entirely at random.

Including the network structure into the bootstrap procedure leads to a broader distribution.





\newpage
\section*{Tables}

  \subsection*{Table 1 --- Evolutionary Analysis of {\it S.cerevisiae}}

    Correlations between evolutionary rate, number of connections and expression level of proteins and the confidence intervals for Kendall's $\tau$ statistic obtained for the different ascomycete species. Values of $\tau$ that have associated p-values $<0.01$ are highlighted in bold. $X1$ denotes correlation with evolutionary rate obtained from a pairwise sequence comparison between \scere\ and species $X$; $X2$ differs from $X1$ only in that the evolutionary rate was obtained using a maximum likelihood estimate. $M$ denotes a rate obtained with respect to {\it S.mikatae}, $B$ to {\it S.bayanus}, $C$ to {\it S.castelii}, $K$ to {\it S.kluyverii}, $A$ to {\it C.albicans}, and $P$ to {\it S.pombe}.\\

    \par

\mbox{
\begin{tabular}{|c|c|c|c|c|c|c|c|c|c|c|c|c|}
\hline 
&\multicolumn{12}{|c|}{Species comparison}\\ \hline
 &M1&M2&B1&B2&C1&C2&K1&K2&A1&A2&P1&P2\\ \hline
{\bf Connectivity}&{\bf -0.13}&{\bf -0.16}&{\bf -0.13}&{\bf -0.14}&{\bf -0.11}&{\bf -0.12}&{\bf -0.12}&{\bf -0.14}&{\bf -0.08}&{\bf -0.08}&{\bf -0.10}&{\bf -0.10}\\
2.5-percentile&-0.17&-0.19&-0.16&-0.17&-0.13&-0.14&-0.15&-0.17&-0.11&-0.11&-0.14&-0.13\\
97.5-percentile&-0.11&-0.13&-0.11&-0.12&-0.08&-0.09&-0.08&-0.10&-0.06&-0.05&-0.07&-0.07\\
\hline{\bf Expression}&{\bf -0.25}&{\bf -0.30}&{\bf -0.26}&{\bf -0.29}&{\bf -0.29}&{\bf -0.32}&{\bf -0.28}&{\bf -0.32}&{\bf -0.28}&{\bf -0.28}&{\bf -0.30}&{\bf -0.28}\\
2.5-percentile&-0.28&-0.33&-0.28&-0.30&-0.31&-0.34&-0.32&-0.35&-0.31&-0.30&-0.33&-0.31\\
97.5-percentile&-0.22&-0.27&-0.24&-0.27&-0.27&-0.30&-0.25&-0.29&-0.26&-0.25&-0.27&-0.25\\
\hline
\end{tabular}
}

\newpage

\subsection*{Table 2 --- Correlations obtained from a direct comparison of \scere\ with \celeg}

Orthologues in the \scere\ and \celeg\ PINs where identified by reciprocal BLAST searches and evolutionary rates, estimated previously (see table 1), were analysed for correlation between evolutionary rate, the number of interactions and expression levels. We also performed an analysis with evolutionary rates estimated directly from the distant \scere\ and \celeg\ comparison.   \\ 
\par    
\mbox{
\begin{tabular}{|c||c|c||c|c|}
\hline
&\multicolumn{2}{c||}{\bf Evolutionary Rate}&\multicolumn{2}{|c|}{\bf Evolutionary Rate}\\
{\bf Comparison}&\multicolumn{2}{c||}{\bf obtained from closely}&\multicolumn{2}{|c|}{\bf obtained from }\\
&\multicolumn{2}{c||}{\bf related species}&\multicolumn{2}{|c|}{\bf \scere-\celeg}\\
\cline{2-5}
&\scere&\celeg&\scere&\celeg\\
\hline
Nr. of Interactions&{\bf -0.11} &{\bf -0.13} &{\bf -0.20} &{\bf -0.10}\\
2.5-percentile&-0.20&-0.23&-0.26&-0.19\\
97.5-percentile&-0.02&-0.04&-0.14&-0.03\\
\hline
Expression& {\bf-0.33}& {\bf -0.44}& {\bf -0.25}&{\bf -0.42} \\
2.5-percentile&-0.41&-0.50&-0.32&-0.47\\
97.5-percentile&-0.24&-0.36&-0.19&-0.37\\\hline
\end{tabular}
}

\subsection*{Table 3 --- Correlations between orthologous proteins in the \scere\ and \celeg\ PINs}

Observed rank correlations (measured by Kendall's $\tau$) for evolutionary rates (measured with respect to {\it S.mikatae} and {\it C.briggsae}, respectively), connectivity and protein expression level (estimated by mRNA expression level in \scere\ and CAI in \celeg).\\

\par 
\mbox{
\begin{tabular}{|c|cc|}
\hline
{\bf Quantity}&{\bf Observed  $\tau$}&{\bf 95\% CI}\\
\hline
 Evolutionary Rate&0.24&0.12-0.35\\
Connectivity&0.07&0.001-0.14\\
Expression&0.32&0.26-0.39\\
\hline
\end{tabular}
}

\newpage
\section*{Supplementary tables}
\subsection*{Table S1 - Estimated evolutionary rates for proteins with different functions}
\par
Averaged estimated evolutionary rates for proteins belonging to each functional class in the gene ontology. \\
\par
\mbox{
\begin{tabular}{|c||c|c|c|c|c|c|c|c|}
\hline
Function&Number&R2&M2&B2&C2&K2&A2&P2\\ \hline
no GO data     &  560           & 0.73& 0.14& 0.15& 0.41& 0.43& 0.57& 0.59\\
chaperone       &  72             & 0.55& 0.07& 0.11& 0.31& 0.36& 0.46& 0.56\\
DNA-binding      & 121             & 0.68& 0.13& 0.15& 0.40& 0.46& 0.60& 0.59\\
enzyme-regulation & 110            & 0.66& 0.12& 0.14& 0.41& 0.40& 0.61& 0.61\\
helicase           &48            & 0.63& 0.09& 0.12& 0.33& 0.33& 0.47& 0.53\\
hydrolase           &278           & 0.61& 0.12& 0.13& 0.34& 0.37& 0.53& 0.59\\
isomerase           &26           & 0.44& 0.07& 0.11& 0.26& 0.27& 0.42& 0.46\\
ligase               &74          & 0.56& 0.12& 0.10& 0.28& 0.28& 0.45& 0.50\\
lyase                &63          & 0.50& 0.06& 0.10& 0.28& 0.27& 0.39& 0.49\\
function unknown     &1484          & 0.72& 0.17& 0.19& 0.47& 0.48& 0.63& 0.68\\
motor activity        & 16        & 0.66& 0.13& 0.18& 0.51& 0.49& 0.48& 0.53\\
nucleotidyltransferase &63        & 0.57& 0.06& 0.10& 0.28& 0.30& 0.46& 0.54\\
oxidoreductase         &158        & 0.45& 0.07& 0.10& 0.29& 0.31& 0.43& 0.49\\
peptidase              &90        & 0.56& 0.08& 0.11& 0.30& 0.31& 0.49& 0.50\\
phosphoprotein-phosphatase&46     & 0.54& 0.08& 0.12& 0.35& 0.37& 0.51& 0.54\\
protein-binding           &216     & 0.70& 0.13& 0.17& 0.41& 0.45& 0.61& 0.62\\
protein-kinase            &107     & 0.61& 0.11& 0.14& 0.38& 0.39& 0.55& 0.63\\
RNA-binding               &178     & 0.68& 0.12& 0.15& 0.37& 0.38& 0.58& 0.60\\
signal-transducer        &56      & 0.65& 0.08& 0.14& 0.34& 0.36& 0.59& 0.62\\
structural-molecule      &211      & 0.62& 0.09& 0.12& 0.35& 0.33& 0.52& 0.58\\
transcription-regulator  &242      & 0.73& 0.13& 0.16& 0.44& 0.44& 0.63& 0.68\\
transferase              &273      & 0.57& 0.11& 0.12& 0.32& 0.37& 0.50& 0.56\\
translation-regulator    &44      & 0.53& 0.06& 0.09& 0.29& 0.28& 0.45& 0.49\\
transporter              &237      & 0.57& 0.11& 0.11& 0.33& 0.35& 0.49& 0.57\\
\hline
\end{tabular}
}

\newpage

\subsection*{Table S2 - Estimated evolutionary rates for proteins involved in different biological processes}
\par
Averaged estimated evolutionary rates for proteins which have been assigned to different biological processes in the gene ontology. \\
\par
\mbox{
\begin{tabular}{|c||c|c|c|c|c|c|c|c|}
\hline
Process&Number&R2&M2&B2&C2&K2&A2&P2\\ \hline
no Go data                  &621          &0.68 &0.14 &0.14 &0.38 &0.37 &0.54 &0.57\\
amino-acid and derivative metabolism &95    &0.48 &0.07 &0.09 &0.23 &0.28 &0.41 &0.48\\
process unknown &1003        &0.69 &0.17 &0.19 &0.47 &0.48 &0.61 &0.67\\
budding              &  24         &0.72 &0.07 &0.18 &0.36 &0.50 &0.55 &0.54\\
carbohydrate-metabolism &89             &0.43 &0.06 &0.10 &0.29 &0.35 &0.42 &0.45\\
cell-cycle                &155         &0.72 &0.15 &0.16 &0.47 &0.51 &0.65 &0.69\\
cell homeostasis          &33           &0.63 &0.12 &0.11 &0.35 &0.41 &0.55 &0.57\\
cellular respiration        &41      &0.59 &0.05 &0.13 &0.36 &0.39 &0.54 &0.56\\
cell-wall organization and biogenesis&87 &0.57 &0.15 &0.15 &0.36 &0.41 &0.56 &0.55\\
coenzyme and prosthetic group metabolism&46&0.55 &0.08 &0.10 &0.32 &0.33 &0.46 &0.55\\
conjugation                    &42&0.74 &0.17 &0.21 &0.50 &0.47 &0.66 &0.74\\
cytokinesis                  &51       &0.71 &0.15 &0.14 &0.38 &0.39 &0.59 &0.66\\
cytoskeleton organization and biogenesis&74&0.75 &0.12 &0.18 &0.41 &0.36 &0.56 &0.56\\
DNA-metabolism                   &204&0.71 &0.11 &0.17 &0.39 &0.43 &0.57 &0.61\\
electron-transport             &6    &0.35 &0.04 &0.06 &0.23 &0.19 &0.36 &0.56\\
energy-pathways                &19     &0.35 &0.09 &0.12 &0.23 &0.35 &0.45 &0.47\\
lipid metabolism           &96        &0.58 &0.08 &0.11 &0.33 &0.34 &0.52 &0.58\\
meiosis              &67          &0.70 &0.22 &0.19 &0.47 &0.48 &0.59 &0.64\\
membrane-organization and biogenesis&14&0.63 &0.11 &0.14 &0.33 &0.43 &0.52 &0.61\\
morphogenesis                   &18  &0.70 &0.11 &0.20 &0.45 &0.47 &0.65 &0.69\\
nuclear-organization and biogenesis &66&0.72 &0.12 &0.18 &0.46 &0.46 &0.66 &0.71\\
organelle-organization and biogenesis&74&0.70 &0.13 &0.15 &0.47 &0.43 &0.61 &0.66\\
protein-biosynthesis          &225 &0.53 &0.08 &0.10 &0.29 &0.31 &0.45 &0.51\\
protein-catabolism         &88       &0.58 &0.12 &0.12 &0.34 &0.32 &0.51 &0.54\\
protein-modification        &214        &0.69 &0.13 &0.14 &0.39 &0.42 &0.57 &0.62\\
pseudohyphal growth   &38   &0.64 &0.15 &0.17 &0.40 &0.43 &0.57 &0.60\\
response to stress    &139            &0.62 &0.11 &0.13 &0.35 &0.39 &0.54 &0.59\\
ribosome-biogenesis and assembly&46  &0.57 &0.08 &0.11 &0.26 &0.29 &0.44 &0.50\\
RNA-metabolism              &274     &0.68 &0.13 &0.15 &0.36 &0.38 &0.55 &0.60\\
signal-transduction         &47      &0.68 &0.13 &0.17 &0.43 &0.42 &0.61 &0.63\\
sporulation              &32     &0.59 &0.15 &0.19 &0.53 &0.48 &0.62 &0.74\\
transcription          &234          &0.70 &0.11 &0.15 &0.40 &0.41 &0.60 &0.62\\
transport                 &372          &0.65 &0.12 &0.14 &0.38 &0.39 &0.56 &0.61\\
vesicle-mediated transport&112       &0.63 &0.14 &0.12 &0.37 &0.40 &0.60 &0.65\\
vitamin-metabolism           &   27  &0.47 &0.07 &0.11 &0.33 &0.31 &0.45 &0.57\\
\hline
\end{tabular}}

\newpage

\subsection*{Table S3 - Estimated evolutionary rates for proteins in different cellular compartments}
\par
Averaged estimated evolutionary rates for proteins with different cellular compartment assignments in the gene-ontology. The is comparatively little variation between compartments and different species comparisons provide qualitatively similar results. \\
\par
\mbox{
\begin{tabular}{|c||c|c|c|c|c|c|c|c|}
\hline
Process&Number&R2&M2&B2&C2&K2&A2&P2\\ \hline
no Go data        &556            &0.72 &0.15 &0.16 &0.40 &0.41 &0.58 &0.59\\
bud                       &64       &0.73 &0.17 &0.19 &0.44 &0.49 &0.64 &0.67\\
cell-cortex             &18    &0.84 &0.10 &0.15 &0.44 &0.50 &0.68 &0.74\\
cellular-component-unknown&478    &0.70 &0.18 &0.22 &0.50 &0.50 &0.63 &0.67\\
cell-wall               &51   &0.55 &0.14 &0.19 &0.38 &0.42 &0.56 &0.62\\
chromosome           &31      &0.71 &0.08 &0.14 &0.39 &0.43 &0.57 &0.62\\
cytoplasm             &841      &0.60 &0.12 &0.13 &0.35 &0.36 &0.52 &0.56\\
cytoplasmic-vesicle     &30     &0.59 &0.15 &0.14 &0.37 &0.44 &0.60 &0.67\\
cytoskeleton           &60    &0.67 &0.11 &0.15 &0.41 &0.41 &0.62 &0.59\\
endo-membrane system    &60    &0.81 &0.16 &0.19 &0.47 &0.43 &0.66 &0.71\\
endoplasmic-reticulum   &192 &0.65 &0.11 &0.13 &0.37 &0.42 &0.56 &0.61\\
extracellular           &10   &0.59 &0.09 &0.13 &0.37 &0.45 &0.53 &0.65\\
Golgi-apparatus        &55    &0.71 &0.19 &0.14 &0.38 &0.41 &0.62 &0.67\\
membrane                  &168  &0.67 &0.15 &0.16 &0.42 &0.40 &0.57 &0.64\\
membrane-fraction        &42  &0.72 &0.17 &0.18 &0.42 &0.40 &0.60 &0.68\\
microtubule-organizing-center&38 &0.80 &0.14 &0.19 &0.57 &0.52 &0.69 &0.66\\
mitochondrial-membrane      &105 &0.58 &0.09 &0.10 &0.36 &0.36 &0.51 &0.57\\
mitochondrion            &248 &0.62 &0.10 &0.14 &0.38 &0.39 &0.53 &0.58\\
nucleolus           &115      &0.62 &0.10 &0.13 &0.33 &0.32 &0.47 &0.56\\
nucleus             &1296        &0.67 &0.13 &0.15 &0.40 &0.42 &0.57 &0.61\\
peroxisome           &22    &0.60 &0.11 &0.17 &0.37 &0.44 &0.55 &0.59\\
plasma-membrane      &125        &0.60 &0.14 &0.15 &0.36 &0.40 &0.55 &0.62\\
ribosome                &118      &0.55 &0.08 &0.10 &0.30 &0.31 &0.49 &0.54\\
site of polarized growth   &5    &0.63 &0.10 &0.21 &0.39 &0.32 &0.59 &0.82\\
vacuole                   &45    &0.67 &0.18 &0.18 &0.41 &0.43 &0.56 &0.64\\
\hline
\end{tabular}
}

\end{bmcformat}
\end{document}